\documentclass{article}
\usepackage{spconf,amsmath,graphicx,hyperref}
\renewcommand{\paragraph}[1]{\vspace{1em}\noindent\textbf{#1}~~}

\hypersetup{hidelinks}

\usepackage{cite}
\usepackage{amsmath,amssymb,amsfonts}
\usepackage{makecell}
\usepackage{tikz}
\usepackage{pgfplots}
\usepgfplotslibrary{statistics}
\pgfplotsset{compat=1.18}
\usepackage{pgf-pie}
\usepackage{tabularx}
\usepackage{algorithm}
\usepackage{algpseudocode}
\usepackage{multirow}
\usepackage{enumitem}
\usepackage{graphicx}
\usepackage{subcaption}
\usepackage{textcomp}
\usepackage{url}

\usepackage[dvipsnames]{xcolor}
\usepackage{booktabs}
\usepackage{fancyhdr}



\begin{document}
\ninept

\title{End-to-End Historical Music Restoration in Latent Space}
%
\name{
Steven Cho$^{1}$
\hspace{0.4em}
Junghyun Koo$^{2}$
\hspace{0.4em}
Raphael Lafargue$^{3}$
\hspace{0.4em}
Tushar Dhyani$^{3}$
\hspace{0.4em}
Eloi Moliner$^{4}$
\hspace{0.4em}
Yuki Mitsufuji$^{2,5}$
}
\address{
$^{1}$ ETH Zurich \qquad
$^{2}$ Sony Corporate Technology Center America, Inc. \\
$^{3}$ Sony Europe Ltd. \quad
$^{4}$ Acoustics Lab, DICE, Aalto University \quad
$^{5}$ Sony Group Corporation
%
}
\maketitle

\begin{abstract}
Historical music restoration (HMR) has almost exclusively focused on constrained problems such as Super-Resolution or the restoration of solo pieces, under-exploring the general task of restoring orchestral historical music, which has multiple instruments. This under-exploration is largely because the HMR domain, early-20th-century recordings, has no pre-degradation ground-truth pairs, making the restoration task unsupervised and more challenging. This paper presents a supervised end-to-end orchestral HMR benchmark by exploring both the synthetic degradation functions and the end-to-end generative deep-learning restoration methods. We simulate the historical recording degradation chain more faithfully than prior work, which makes orchestral restoration into a tractable supervised problem. A latent flow-matching model trained on the resulting synthetic pairs outperforms existing HMR baselines on intrusive, non-intrusive, and subjective evaluations. We also curate and release a 9.3-hour license-free, unpaired, historical classical-music test set, along with code and audio demos \footnote{\url{https://github.com/SonyResearch/historical_music_restoration}}.
\end{abstract}

\begin{keywords}
Audio, Restoration, Dataset, Machine Learning, Music.
\end{keywords}

\section{Introduction}

In 1877, Thomas Edison invented the phonograph, giving the world, for the first time, the ability to record music \cite{edison1880phonograph}. Then in only 100 years, recording technology rapidly advanced to almost lossless recordings. Still, many of the time's legendary artists such as Fats Waller, Tchaikovsky, and Debussy recorded a vast amount of their discography with lossy and noisy recording equipment such as wax cylinders, 78-rpm shellac disks, and gramophones, leaving much of their valuable works corrupted and underappreciated. This makes the historical music restoration task valuable for the rediscovery and appreciation of foundational works of many genres. For this paper, we define music published before 1926 as historical for electrical recording was popularized in 1925, rapidly improving recording quality after.

Historical music restoration began with classical digital signal processing methods, including blind equalization using spectral averages \cite{stockham1975blind} and statistical restoration methods for clicks and hiss suppression ~\cite{godsill1998digital}.
Eventually, neural systems trained on pairs generated by synthetically adding noise to clean music were shown to denoise historical recordings \cite{li2020denoise, moliner2022two}.
Pair-trained neural systems also proved effective in restoration related domain of Super-Resolution (SR) - the task of recovering high-frequency information lost to ideal low-pass filters \cite{liu2023audiosr}. SR methods successfully applied generative methods such as flow matching \cite{lipman2022cfm} or Schrodinger Bridges \cite{kong2025a2sb} on paired waveform domains \cite{Lee_2021} and frequency domain representations \cite{moliner2022behm, yun2025flowhigh}. 

While the waveform and time-frequency space have been adequate for training speech restoration models, training a generalizable restoration model on music is much more difficult due to its larger frequency range and spectral diversity \cite{im2026saga}. Therefore, newer models began applying predictive methodologies onto more compact and relevant features - latents - of waveforms and spectrograms by passing them through domain-aligned neural codecs before restoration \cite{liu2023audiosr, liu2026latentflowsr, im2026saga}.
\begin{figure}[t]
    \centering
    \includegraphics[width=\columnwidth]{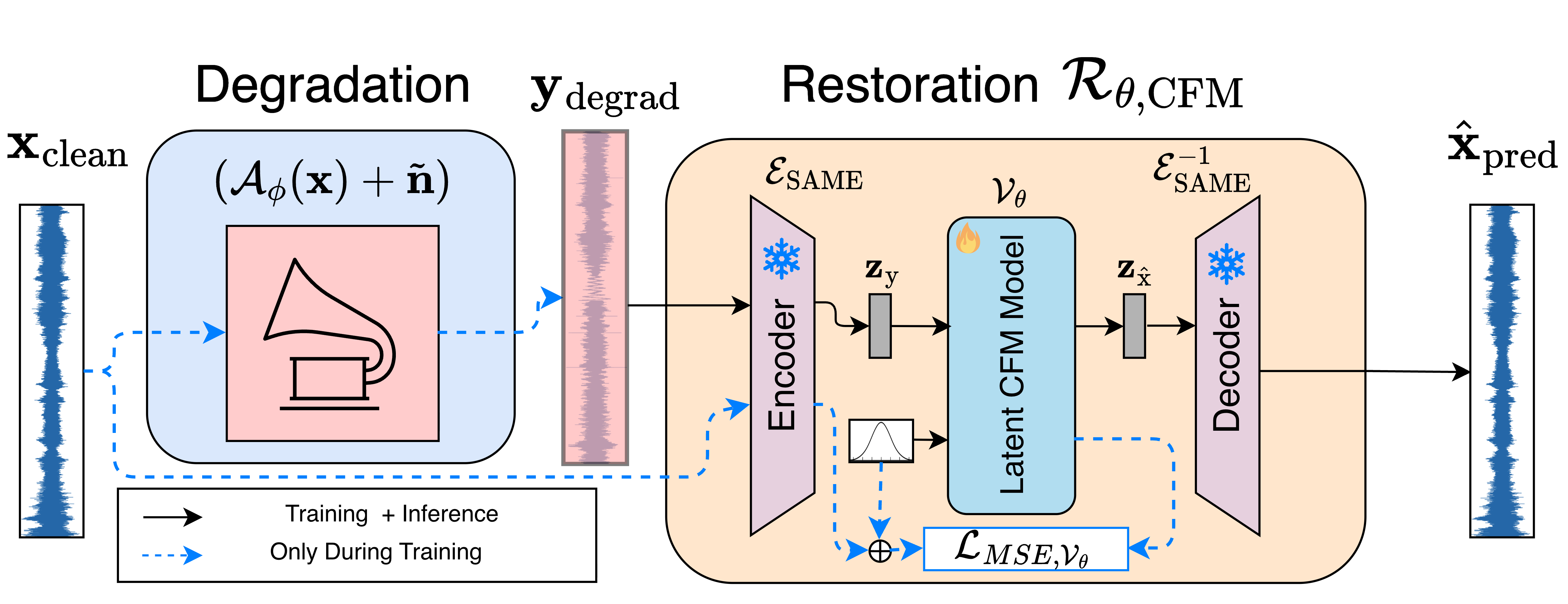}
    \caption{
    Overview of the proposed historical music restoration pipeline $\mathcal{R}_{\theta,\mathrm{CFM}}$.
    }
    \label{fig:restoration_pipeline}
    \vspace{-6mm}
\end{figure}
While this significantly improved music SR models' generalization \cite{im2026saga}, because historical music degradations never exclusively involve ideal low-pass filters, the SR models still fail to restore historical music due to its recording-dependent nonlinear distortions and spectral attenuations \cite{moliner2024blind}. As a result, some methods opt to restore in the unpaired blind inverse problem space in the frequency domain, blindly inferring the degradation of the audio while simultaneously restoring it \cite{moliner2024blind, moliner2024diffusion}. Still, the methodology has only been proven for single piece music such as piano and vocals \cite{moliner2024diffusion}.

Therefore, we propose a synthetic supervision approach. By introducing a degradation function with stochastic distortions covering bandwidth limitation and spectral coloration of previous works \cite{moliner2022behm, moliner2024diffusion} and gramophone noise and nonlinear system behavior, it can sufficiently model the early 20th century recorded music from modern high-fidelity recordings. The synthetically degraded recordings supervise a simple end-to-end latent conditional flow matching model to outperform all previous  historical restoration models, providing a strong baseline for the orchestral historical music restoration space.

To introduce the problem space of orchestral historical music restoration, we make the following contributions:

\begin{itemize}[leftmargin=1em,labelsep=0.4em]
\item We introduce an \textbf{unpaired 149-song test set} of real historical orchestral recordings for evaluating restoration.
\item We propose a \textbf{5-stage synthetic degradation function} based on a Wiener–Hammerstein structure 
to synthesize paired training data.  
\item Using the degradation function, we apply a \textbf{Latent-space Conditional Flow Matching (CFM) restoration model} to orchestral historical music restoration, substantially outperforming prior music restoration baselines.
\end{itemize}


\section{Methods}
\subsection{Problem Definition}

Let $\mathbf{y} \in \mathbb{R}^{T}$ denote an observed orchestral historical music recording of duration $T$ samples. We assume this recording originates from an unknown, clean, high-fidelity recording of the same duration, denoted $\mathbf{x} \in \mathbb{R}^T$. The relationship between the two is given by the forward expression
\begin{equation}
\mathbf{y}=\mathcal{A}(\mathbf{x})+\mathbf{n},
\end{equation}
where $\mathbf{n}$ represents additive gramophone degradations in historical recordings, such as hiss and clicks \cite{godsill1998digital}, and $\mathcal{A}$ denotes the composition of non-additive degradations, including bandwidth limitation, spectral coloration, nonlinear distortion, and other unknown degradation processes.

Historical music restoration therefore seeks to recover an estimate $\hat{\mathbf{x}} \approx \mathbf{x}$ of the unknown high-fidelity recording from $\mathbf{y}$ alone.
Critically, both $\mathcal{A}$ and $\mathbf{n}$ are unknown, making historical music restoration a blind inverse problem: the degradation process must be inverted without any explicit knowledge of what needs to be corrected. 
This blindness renders the problem severely ill-posed as infinitely many pairs $(\mathbf{x}, \mathcal{A})$ could produce $\mathbf{y}$. Therefore, recovering $\hat{\mathbf{x}}$ requires a strong prior of plausible high-fidelity music and the degradation process itself. Hence, the synthetic supervision attempts to provide both.

\subsection{Synthetic Degradation Function}
\label{sec: synth_degrad_model}

First, rather than attempting to match the true distribution of real degradations $\mathcal{A}+\mathbf{n}$, we design $\mathcal{A}_\phi + \tilde{\mathbf{n}}$, whose parameters $\phi$ can be randomly sampled during training. $\mathcal{A}_\phi + \tilde{\mathbf{n}}$, together with parameter distribution $p_\phi$, induces a distribution over operators broad enough to encompass $\mathcal{A} + {\mathbf{n}}$. Training on examples $\mathcal{A}_\phi(\mathbf{x})+\tilde{\mathbf{n}}$ with $\phi \sim p_\phi$ then exposes the restoration model $\mathcal{R}_\theta$ to a wider family of degradations than it will encounter at inference, encouraging generalization to real recordings $\mathbf{y} = \mathcal{A}(\mathbf{x}) + \mathbf{n}$ without requiring $\mathcal{A}$ to be identified.

We build $\mathcal{A}_\phi$ around a Wiener--Hammerstein (WH) model classically used for nonlinear system identification \cite{schoukens2019nonlinear}, consisting of a static nonlinearity $\Psi$ placed between two linear spectral filters $H_\mathrm{pre}$ and $H_\mathrm{post}$. 
This structure provides a compact approximation of historical recording and playback chains, in which frequency-dependent coloration interacts with amplitude-dependent distortion.
The complete function is given by
\begin{equation}
\mathcal{A}_\phi = H_\mathrm{BPF} \circ H_\mathrm{post} \circ \Psi_\mathrm{nonlin} \circ H_\mathrm{pre}.
\end{equation}

The linear blocks $H_\mathrm{pre}$ and $H_\mathrm{post}$ are zero-phase finite impulse response (FIR) filters applied in the frequency domain, similar to \cite{colonel2021reverse, moliner2025unsuperviseddafx}. Their magnitude responses are parameterized by a set of attenuation weights ($\phi_\mathrm{pre}$ and $\phi_\mathrm{post}$), one per octave band, which are linearly interpolated, while the phase response is constrained to zero.
An additional bandpass filter $H_\mathrm{BPF}$ complements $H_\mathrm{post}$ and explicitly models the restricted bandwidth typical of historical recordings. It is likewise a zero-phase FIR filter implemented in the frequency domain, but follows the parameterization of \cite{moliner2024diffusion}, comprising low and high cutoff frequencies ($f_\mathrm{low}$ and $f_\mathrm{high}$) together with two roll-off slopes ($s_\mathrm{low}$ and $s_\mathrm{high}$) that define the attenuation rate beyond each cutoff.

The nonlinear block $\Psi_\mathrm{nonlin}$ is a static nonlinearity, implemented as a scaled $\tanh$ \cite{colonel2021reverse} with a wet/dry residual connection,
\begin{equation}
\Psi_\mathrm{nonlin}(\mathbf{x})=(1-w)\mathbf{x}+w\tanh(a\mathbf{x})/a,
\end{equation}
with parameters $\phi_\mathrm{nonlin}=\{w, a\}$. The additive degradation $\tilde{\mathbf{n}}$ \cite{moliner2022two} signal-to-noise ratio is set by $\phi_\mathrm{add}$.

The full parameter set $\phi=\{\phi_{\mathrm{pre}}, \phi_{nonlin}, \phi_{\mathrm{post}},\phi_\mathrm{BPF}, \phi_{\mathrm{add}}\}$ collects the coefficients of all four blocks plus additive gramophone noise and is resampled independently for every training example. The choice of $p_\phi$ is critical: distributions that are too narrow risk teaching $\mathcal{R}_\theta$ to invert one particular degradation, while deliberately broad distributions favor robustness to unknown historical conditions at the cost of some realism. We designed $p_\phi$ by analyzing a dataset of historical recordings and through qualitative listening, and report the resulting parameters in the official repository. 

\begin{table}[t]
\centering
\caption{\textbf{MAD ($\downarrow$) between degradation ablations embeddings on FOS and historical unpaired test set.}
The final row directly compares the two subsets of the unpaired test set.}
\vspace{-3mm}
\label{tab:mad-scores-FOS-degradation}

\scriptsize
\renewcommand{\arraystretch}{1.18}
\setlength{\tabcolsep}{1.5pt}

\begin{tabular*}{\columnwidth}{
@{\extracolsep{\fill}}
l
cc
cc
cc
@{}
}
\toprule


& \multicolumn{2}{c}{\textbf{MERT~\cite{huang2025aligning}}}
& \multicolumn{2}{c}{\textbf{Fx-Enc.~\cite{yeh2025fxencoder}}}
& \multicolumn{2}{c}{\textbf{SAME-L~\cite{parker2026same}}} \\

\cmidrule(lr){2-3}
\cmidrule(lr){4-5}
\cmidrule(lr){6-7}

\textbf{Degradation}
& \makecell{\textbf{Full}\\\textbf{Orch.}}
& \makecell{\textbf{Light}\\\textbf{Orch.}}
& \makecell{\textbf{Full}\\\textbf{Orch.}}
& \makecell{\textbf{Light}\\\textbf{Orch.}}
& \makecell{\textbf{Full}\\\textbf{Orch.}}
& \makecell{\textbf{Light}\\\textbf{Orch.}} \\

\midrule

Clean
& 5.39
& 5.40
& 5.43
& 5.49
& 5.33
& 5.45 \\

Gaussian Noise
& 5.18
& 5.13
& 4.71
& 5.08
& 5.40
& 5.44 \\

Low-Pass + Gaussian Noise
& 5.31
& 5.46
& 5.42
& 5.49
& 5.24
& 5.37 \\

Gramophone Noise
& \textbf{3.90}
& \textbf{4.42}
& \textbf{3.86}
& \textbf{4.63}
& \underline{4.57}
& \textbf{4.82} \\

WH + Gramophone Noise (Ours)
& \underline{4.23}
& \underline{4.55}
& \underline{4.42}
& \underline{4.99}
& \textbf{4.54}
& \underline{4.95} \\

\midrule

\textbf{Full-Orch. vs. Light-Orch.}
& \multicolumn{2}{c}{2.08}
& \multicolumn{2}{c}{2.67}
& \multicolumn{2}{c}{3.00} \\

\bottomrule
\end{tabular*}
\vspace{-0.4cm}
\end{table}
We then validated our degradation function by computing the MAD \cite{huang2025aligning} between the synthetically degraded output with the unpaired test set, and our degradation function significantly outperformed other common functions except for 'Gramophone Only' as shown in Table \ref{tab:mad-scores-FOS-degradation}. The Gramophone Noise method is ablated and analyzed in Section \ref{sec: experimental-setup-subject} and \ref{sec: results}, and our proposed 5-stage degradation function still restores best. MAD was chosen over Frechet Audio Distance 
for its improved human music perceptual correlation with audio degradations and musical structure \cite{huang2025aligning}.

\subsection{Restoration Model}

With our synthetic pairs $(\mathbf{x}, \mathcal{A}_\phi(\mathbf{x}) + \tilde{\mathbf{n}})$ from the degradation function, we adopt a supervised-learning paradigm, which has proven effective across a wide range of blind inverse problems in audio and image restoration \cite{moliner2022behm, saharia2021image}. We train a restoration model $\mathcal{R}_\theta: \mathbf{y} \mapsto \hat{\mathbf{x}}$, parametrized by $\theta$, that approximates the inverse of the degradation process with paired examples of degraded and clean signals $(\mathbf{x}, \mathbf{y})$. The restoration pipeline overview is shown in Figure \ref{fig:restoration_pipeline}.

Latent-space generative modeling has demonstrated promising results in supervised audio SR \cite{liu2026latentflowsr, liu2023audiosr}, therefore before any restoration we employ an autoencoder $\mathcal{E}: \mathbb{R}^T \rightarrow \mathbb{R}^{C\times N}$ to first embed the historical recordings into latent vectors $\mathbf{z}_y\in \mathbb{R}^{C\times N}$ where $C$ and $N$ are the number of channels and frames respectively. This reduces the restoration problem from $\mathbf{y} \mapsto \mathbf{x}$ to $\mathbf{z_y} \mapsto \mathbf{z_x}$.

Then a latent-space conditional flow matching restoration model $\mathcal{R}_{\theta,\mathrm{CFM}}$, inspired by Liu et al.'s LatentFlowSR \cite{liu2026latentflowsr}, is applied but without additional spectral and semantic embeddings. The training, inference, and objective is based on Lipman et al.'s Flow matching configuration \cite{lipman2022cfm}:
\begin{equation}
\mathcal{L}_{\mathrm{CFM}}
=
\mathbb{E}_{\mathbf z_0,t,\mathbf{x},\phi}
\left[
\left\|
\mathcal V_\theta(\mathbf z_t,t,\mathcal{E}(\mathcal{A}_\phi(\mathbf{x})+\tilde{\mathbf{n}}))-\mathbf v
\right\|_2^2
\right],
\end{equation}
where
$\mathbf{z}_0\sim\mathcal{N}(\mathbf{0},\mathbf{I})$,
$t\sim\mathcal U(0,1)$,
$\mathbf x\sim p_{\mathbf{x}}$,
and $\phi\sim p_\phi$. 
Defining $\mathbf{z}_{\mathbf{x}} = \mathcal{E}(\mathbf{x})$, 
the latent condition is derived with $\mathbf{z}_{\mathrm{t}} = (1-t)\mathbf{z_0} + t\mathbf{z_\mathbf{x}}$, and the target velocity is defined as $\mathbf{v} = (\mathbf{z}_\mathbf{x}-\mathbf{z}_0)$.
During training, a time-conditioned transformer learns the velocity $\mathcal{V}_{\theta}$ from the Gaussian source toward the clean target $\mathbf{v}$ conditioned with $\mathbf{z}_y =\mathcal{E}(\mathcal{A}_\phi(\mathbf{x}) + \tilde{\mathbf{n}})$.
During inference, the historical music input $\mathbf{y}$ is instead used to create the $\mathbf{z}_{\mathrm{y}}$ condition, and the learned velocity field is integrated using the Euler ODE solver \cite{lipman2022cfm}.

\section{Experimental Setup}








\begin{table}[t]
\centering
\caption{\textbf{Autoencoder reconstruction Mel-MSE with equal-window aggregation.}
Waveforms were encoded and decoded by the corresponding autoencoder and compared against the input using Mel-MSE}
\label{tab:codec-reconstruction-mel-mse-window}
\vspace{-2mm}

\small
\renewcommand{\arraystretch}{1.15}
\setlength{\tabcolsep}{4pt}

\resizebox{\columnwidth}{!}{%
\begin{tabular}{@{}lcccc@{}}
\toprule
\textbf{Codec}
& \makecell{\textbf{Paired}\\\textbf{Clean}}
& \makecell{\textbf{Paired}\\\textbf{5-Stage}}
& \makecell{\textbf{Full}\\\textbf{Orchestra}}
& \makecell{\textbf{Light}\\\textbf{Orchestra}} \\
\midrule

CoDiCodec \cite{pasini2025codicodec}
& $1.08 \pm 0.45$
& $0.25 \pm 0.27$
& $1.37 \pm 0.68$
& $1.24 \pm 0.69$ \\

SAO~\cite{evans2025stable}
& $0.39 \pm 0.19$
& $0.12 \pm 0.17$
& $0.80 \pm 0.50$
& $0.88 \pm 0.58$ \\

DAC \cite{kumar2023high}
& $\underline{0.22 \pm 0.10}$
& $\underline{0.08 \pm 0.15}$
& $\underline{0.72 \pm 0.57}$
& $\underline{0.87 \pm 0.63}$ \\

SAME-L
& $\mathbf{0.11 \pm 0.08}$
& $\mathbf{0.05 \pm 0.12}$
& $\mathbf{0.62 \pm 0.57}$
& $\mathbf{0.81 \pm 0.62}$ \\

\bottomrule
\end{tabular}%
}
\vspace{-4mm}
\end{table}

\label{sec:experimental-setup}

\subsection{Autoencoder and Backbone Architecture}

For the encoder $\mathcal{E}$ and decoder $\mathcal{E}^{-1}$, we use SAME-L the autoencoder used in Stable Audio 3 \cite{parker2026same}.
This choice was made through preliminary experiments that gauged multiple SOTA neural codecs' degraded music reconstruction capabilities. See Table \ref{tab:codec-reconstruction-mel-mse-window}.

The backbone for $\mathcal{V}_{\theta}$ contains approximately 40M parameters, and consists of a transformer architecture: 576-dimensional tokens, 8 AdaLN-Zero blocks, 9 attention heads, and an MLP width of 1152. The CFM was trained for 400k optimizer steps using AdamW with a learning rate of \(2\times10^{-4}\) and a global batch size of 768 on four NVIDIA A100 GPUs. At inference, we use 10 Euler steps.

\begin{table*}[t]
\centering
\caption{\textbf{Restoration performance on objective and subjective evaluations.}}
\label{tab:FOS-restoration-evaluation}
\vspace{-2mm}
\scriptsize
\renewcommand{\arraystretch}{1.15}
\setlength{\tabcolsep}{1.5pt}
\begin{tabular*}{\textwidth}{@{\extracolsep{\fill}}l c cccc cccc@{}}
\toprule
\multirow{3}{*}{\textbf{Method}}
& \multirow{3}{*}{\textbf{Params}}
& \multicolumn{4}{c}{\textbf{Historical Unpaired Test Set}}
& \multicolumn{4}{c}{\textbf{Synthetic Paired Test Set}} \\
\cmidrule(lr){3-6}\cmidrule(lr){7-10}
& & \multicolumn{2}{c}{\textbf{AA-PQ} ($\uparrow$)}
& \multirow{2}{*}{\makecell{\textbf{MOS-P}\\($\uparrow$)}}
& \multirow{2}{*}{\makecell{\textbf{MOS-Q}\\($\uparrow$)}}
& \multirow{2}{*}{\textbf{AA-PQ} ($\uparrow$)}
& \multirow{2}{*}{\makecell{\textbf{Mel}\\\textbf{MSE} ($\downarrow$)}}
& \multirow{2}{*}{\makecell{\textbf{CLAP}\\\textbf{Cos.} ($\uparrow$)}}
& \multirow{2}{*}{\makecell{\textbf{MAD}\\($\downarrow$)}} \\
\cmidrule(lr){3-4}
& & \makecell{\textbf{Full-Orch.}} & \makecell{\textbf{Light-Orch.}} & & & & & & \\
\midrule
Unprocessed input & -- & 4.94 & 5.00 & -- & 1.17 & 5.74 & 3.94 & 0.44 & 5.03 \\
Ground truth & -- & -- & -- & -- & 4.44 & 7.44 & 0.00 & 1.00 & 0.00 \\
\midrule
BEHM-GAN & 82M & 4.87 & 4.93 & -- & -- & 6.05 & 5.33 & 0.45 & 4.94 \\
BABE2 & 40M & 5.88 & 5.87 & 3.70 & \underline{3.52} & 6.35 & 4.99 & 0.56 & 5.46 \\
Denoiser & 73M & 5.74 & 6.02 & -- & -- & 6.35 & 5.27 & 0.51 & \textbf{1.21} \\
\midrule
BEHM-GAN-FOS & 82M & 5.67 & 5.66 & -- & -- & 6.09 & \underline{1.97} & 0.65 & 4.09 \\
BABE2-FOS & 40M & 5.54 & 5.41 & 3.90 & 2.72 & 6.24 & 3.22 & 0.65 & \underline{1.82} \\
Denoiser-FOS & 73M & 5.95 & \underline{6.17} & -- & -- & \underline{6.88} & 2.25 & \underline{0.71} & 4.19 \\
\addlinespace[2pt]
\textbf{SAMECFM} & 40M & \textbf{6.70} & \textbf{6.86} & \textbf{4.32} & \textbf{3.89} & \textbf{7.21} & \textbf{1.90} & \textbf{0.73} & 3.70 \\
\quad SAMECFM-G & 40M & \underline{6.07} & 6.06 & \underline{4.26} & 3.22 & 6.08 & 3.92 & 0.54 & 5.10 \\
\bottomrule
\end{tabular*}
\vspace{-2mm}
\end{table*}

\subsection{Training and Validation Data}
\label{sec:public-training-data}

 We restrict our historical music restoration to classical music. Due to abundance of high quality public classical music datasets and genre-inherent consistency of piece renditions, this choice minimizes genre-related out of distribution concerns when generating degraded-clean paired training sets to model real historical recordings. The paper does not address music with vocals as vocal restoration is an independent restoration field. 

To incorporate the training data pairs $(\mathbf{x}, \mathcal{A}_\phi(\mathbf{x}) + \tilde{\mathbf{n}})$, we construct a public instrumental-classical corpus from PHENICX-Anechoic~\cite{miron2016phenicx},
ChoraleBricks~\cite{balke2025choralebricks},
URMP~\cite{li2018creating}, ORCHSET~\cite{bosch2016orchset}, and the Beethoven Symphony Dataset (BSD)~\cite{berendes2026beethoven} for $\mathbf{x}$. The resulting training corpus \emph{Full-Orchestra + sections} (FOS) contains 5.17 hours of Full-Orchestra music, with 1.65 hours additional brass, woodwind, and string section recordings from PHENICX-Anechoic, ChoraleBricks, and URMP, totaling approximately 6.82 hours of classical music data: Full-Orchestra recordings and section recordings account for 75.8\% and 24.2\%, respectively. All audio is preprocessed to mono, resampled to 44.1\,kHz, and loudness normalized to $-23$\,LUFS. Full-Orchestra describes a symphonic orchestra recording.
 
Section mixtures are included to increase available
training data while directly exposing instrument-family specific timbre and high-frequency structure to the model. Finally, FOS corpus was segmented into 5-s windows using a 10\% window hop size for training.

For the validation set, validation windows
are randomly assigned with overlap-aware data leakage prevention: once a window is assigned for validation, all windows overlapping that window are discarded. Random validation window assignments occur until the train-to-validate window ratio reaches 9:1.

For each training and validation window, 20 iterations of $\mathcal{A}_\phi(\mathbf{x})+\tilde{\mathbf{n}}$ are applied, encoded, and saved along with the ground truth for efficient training. The synthetic degradation function is applied only after the train-validation partitioning to ensure differently degraded versions of the same
window cannot leak into the training set. 
\subsection{Test Data}
\label{sec:test-data}

We curate a paired test set and an unpaired historical test set. The paired set is derived from the Spheres classical music dataset (53.0\,min) \cite{garcia2026spheres}. We only use their provided 'StereoMix' Full-Orchestra recordings. All audio is preprocessed like the training set except the recordings are divided into non-overlapping 5-s windows.

The unpaired historical music test set (9.3\,h) consists of digitized recordings collected from Internet Archive and is exclusively instrumental orchestra classical music recorded and published in the United States between 1911 and 1925. It is also upsampled to 44.1 kHz FLAC files, converted into mono, and loudness-normalized to $-23$ LUFS. In addition, it is sorted into two groups: Full-Orchestra and Light-Orchestra -- wind-ensemble and chamber recordings. This will serve to analyze the importance of sub-genre train-test domain matching for historical restoration. To the best of our knowledge, this is the first test set curated for the historical music restoration space. We release the official unpaired test data to serve as a benchmark for future research works. \footnote{\url{https://zenodo.org/records/22737610}}

\pgfplotsset{compat=1.18}
\subsection{Objective Evaluation}
\label{subsec:evaluation-protocol}

Each method restores every window in both the paired and unpaired test sets. Metric scores are first averaged with equal weight across windows within each song and then across songs to obtain a single test-set score, reported in Table \ref{tab:FOS-restoration-evaluation}. If needed, outputs are resampled before scoring (e.g.\ 22.05\,kHz BEHM-GAN) for equal comparisons with methods operating at 44.1\,kHz. For historical unpaired test set, only reference-free metrics were calculated on both Full-Orchestra and Light-Orchestra. 

AA-PQ (Audiobox Aesthetic Production Quality) \cite{tjandra2025meta} is an established reference-free objective metric of perceived production quality sensitive to frequencies and dynamics. Mel-MSE \cite{kong2020hifigan} is the mean-squared-error between mel-spectrograms of the restored
and reference audio, providing a perceptually weighted measure of tonal and timbral reconstruction quality. CLAP Cos. \cite{elizalde2023clap} measures a pair's cosine similarity in CLAP's musical/semantic embedding space, which encodes instrumentation, musical character, timbral identity, and broad acoustic characteristics. Finally, MAD \cite{huang2025aligning} measures the divergence between the restored and ground-truth test audio distributions in the MERT embedding space.

Taken together, the four metrics  jointly cover \emph{spectral fidelity} and \emph{content preservation} -- both essential when restoring historical recordings with minimal hallucinations and high production quality. They also validate the reference-free metrics on the unpaired test set.

\subsection{Subjective Evaluation}
\label{sec: experimental-setup-subject}

\begin{figure}[t]
\centering
\begingroup
\newlength{\mosplotheight}
\setlength{\mosplotheight}{2.75cm}

\begin{minipage}[t]{0.5\columnwidth}
\centering
\begin{tikzpicture}
\begin{axis}[
  title={\footnotesize\textbf{MOS-Quality}},
  title style={at={(axis description cs:0.5,1.03)},anchor=south},
  width=0.74\linewidth,
  height=\mosplotheight,
  scale only axis,
  ymin=1,
  ymax=5,
  ytick={1,2,3,4,5},
  yticklabels={Bad 1,Poor 2,Fair 3,Good 4,Excel. 5},
  xtick={1,2,3,4,5,6},
  xticklabels={Input,BABE2-P,BABE2-F,SAMECFM-G,SAMECFM,GT},
  xticklabel style={
    font=\tiny,
    rotate=35,
    anchor=north east,
    inner sep=0pt,
    yshift=-1mm
  },
  grid=major,
  grid style={gray!20},
  boxplot/draw direction=y,
  boxplot={box extend=0.35},
  boxplot/every box/.style={thin},
  boxplot/every median/.style={black,semithick},
  boxplot/every whisker/.style={thin}
]
\addplot+[draw=blue!75!black,fill=blue!12,boxplot prepared={draw position=1,lower whisker=1.000,lower quartile=1.055,median=1.170,upper quartile=1.286,upper whisker=1.500}] coordinates {};
\addplot+[draw=red!75!black,fill=red!12,boxplot prepared={draw position=2,lower whisker=2.250,lower quartile=3.243,median=3.523,upper quartile=3.803,upper whisker=4.750}] coordinates {};
\addplot+[draw=orange!85!black,fill=orange!18,boxplot prepared={draw position=3,lower whisker=2.000,lower quartile=2.473,median=2.716,upper quartile=2.959,upper whisker=3.750}] coordinates {};
\addplot+[draw=green!55!black,fill=green!18,boxplot prepared={draw position=4,lower whisker=2.000,lower quartile=2.966,median=3.216,upper quartile=3.466,upper whisker=4.000}] coordinates {};
\addplot+[draw=violet!75!black,fill=violet!15,boxplot prepared={draw position=5,lower whisker=3.250,lower quartile=3.706,median=3.886,upper quartile=4.066,upper whisker=4.750}] coordinates {};
\addplot+[draw=cyan!70!black,fill=cyan!15,boxplot prepared={draw position=6,lower whisker=4.000,lower quartile=4.298,median=4.443,upper quartile=4.588,upper whisker=5.000}] coordinates {};
\end{axis}
\end{tikzpicture}
\end{minipage}%
\hfill
\begin{minipage}[t]{0.5\columnwidth}
\centering
\begin{tikzpicture}
\begin{axis}[
  title={\footnotesize\textbf{MOS-Preservation}},
  title style={at={(axis description cs:0.5,1.03)},anchor=south},
  width=0.82\linewidth,height=\mosplotheight,scale only axis,
  ymin=1,ymax=5,ytick={1,2,3,4,5},yticklabels={},
  yticklabel style={font=\tiny},xtick={1,2,3,4},
  xticklabels={BABE2-P,BABE2-F,SAMECFM-G,SAMECFM},
  xticklabel style={
    font=\tiny,
    rotate=35,
    anchor=north east,
    inner sep=0pt,
    yshift=-1mm
  },
  grid=major,grid style={gray!20},boxplot/draw direction=y,
  boxplot={box extend=0.18},boxplot/every box/.style={thin},
  boxplot/every median/.style={black,semithick},boxplot/every whisker/.style={thin}
]
\addplot+[draw=red!75!black,fill=red!12,boxplot prepared={draw position=1,lower whisker=2.500,lower quartile=3.340,median=3.705,upper quartile=4.069,upper whisker=4.750}] coordinates {};
\addplot+[draw=orange!85!black,fill=orange!18,boxplot prepared={draw position=2,lower whisker=2.500,lower quartile=3.616,median=3.898,upper quartile=4.179,upper whisker=5.000}] coordinates {};
\addplot+[draw=green!55!black,fill=green!18,boxplot prepared={draw position=3,lower whisker=3.000,lower quartile=3.995,median=4.261,upper quartile=4.527,upper whisker=5.000}] coordinates {};
\addplot+[draw=violet!75!black,fill=violet!15,boxplot prepared={draw position=4,lower whisker=3.000,lower quartile=4.057,median=4.318,upper quartile=4.579,upper whisker=5.000}] coordinates {};

\end{axis}
\end{tikzpicture}
\end{minipage}

\endgroup
\vspace{-3mm}

\caption{MOS-Q and MOS-P means, 95\% confidence intervals, and non-outlier ranges}
\label{fig:subjective-mos-overall}
\vspace{-4mm}
\end{figure}
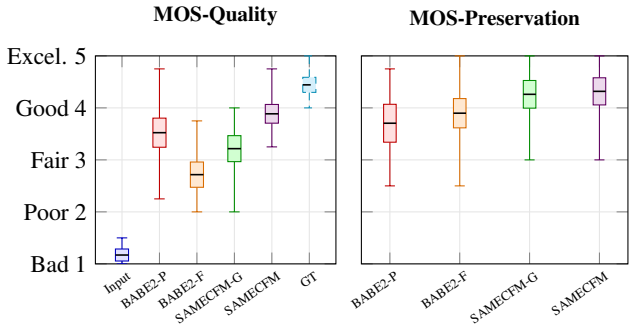

For subjective evaluation, we conduct MOS tests on two different categories: Quality (MOS-Q) and Preservation (MOS-P) of historical music. MOS-Q evaluates overall audio quality across low/high anchors, our methods (SAMECFM $\mathcal{R}_{\theta,\mathrm{CFM}}$ and SAMECFM-G), and BABE2 methods - BABE2 was chosen as it is the most recent high-performing generative restoration method in historical music restoration. See Figure \ref{fig:subjective-mos-overall}. The difference between SAMECFM and SAMECFM-G is that the paired FOS train set for the latter only degrades with the gramophone noise  $\tilde{\mathbf{n}}$ as ablation. Listeners rated the audio quality on a 1--5 scale across noise, timbre, clarity, and high-frequency details. MOS-P evaluates how well four restoration methods -- our  methods and BABE2 methods -- preserve the historical input and avoid hallucinations. Listeners were asked how well on a 1--5 scale the restored audio retains the essential features of the degraded input audio, regarding melody, instrumentation, and style. Both tests use ten 5-s excerpts equally divided between Full-Orchestra and Light-Orchestra recordings. 38 listeners were asked, and 22 were valid. We removed a total of 16 testers due to failing to rate the high anchor ground truth with an average score of at least 4 and the low anchor input with an average score of less than 2. Each method received 88 ratings on both MOS tests. We report mean scores with 95\% confidence intervals. 
\label{subsec:evaluation-subjective}

\section{Results}
\label{sec: results}

We compared pretrained and FOS-domain-adapted BEHM-GAN, BABE2, and Denoiser
against our proposed SAMECFM. Domain adaptation
uses the FOS dataset together with our proposed historical degradation function.

From the experiments in Table \ref{tab:FOS-restoration-evaluation}, both objective and subjective metrics have shown that our WH + Gram. Noise  degradation function supervises better restoration than the Gram. Noise Only ablation. Gram. Noise Only achieved a lower MAD in Table \ref{tab:mad-scores-FOS-degradation} than WH + Gram. Noise likely because MERT, Fx-Encoder++, and SAME-L were not intended to embed historical degradation. Still, SAME-L, due to its autoencoder reconstruction training, possibly encoded more degradation related information, leading to WH + Gram. Noise's one win over Gram. Noise.

Degradation matching is also shown to be more important than sub-genre matching. In Table~\ref{tab:FOS-restoration-evaluation}, according to AA-PQ, the pretrained baselines generally see greater improvements from unpaired set to paired set than Light-Orchestra to Full-Orchestra, where the latter of both is better domain aligned. Their FOS adaptations and methods, in addition to improving all metrics, follow the same behavior. The larger impact of unpaired to paired shift over sub-genre shift demonstrates the importance of accurate target degradation simulation over sub-genre alignment. 

Overall, SAMECFM achieved the best unpaired metric, MOS-Quality and MOS-Preservation scores, despite the lowest parameter count, highlighting the strength of  effective degradation functions and supervised methods in the historical degradation restoration domain. 
A notable trade-off is generative methods slightly sacrificed preservation for quality. Although SAMECFM and SAMECFM-G received the highest MOS-Preservation scores, their MAD on paired test set ranked third and seventh respectively, suggesting some hallucination. 
Finally, among all the objective metrics, we found that AA-PQ best aligned with MOS-Q with Pearson correlations of $r=0.920$ for Full-Orchestra and $r=0.867$ for Light-Orchestra, supporting its use as a reference-free quality metric. 

\section{Conclusion}

This paper introduces End-to-End Historical Music Restoration as a blind inverse problem extending beyond conventional super-resolution and single-instrument restoration. Across objective and human evaluations, domain adaptation using Full-Orchestra and section-level training data, together with a broader nonlinear historical degradation function, substantially improves restoration quality.
Our latent CFM model achieves the strongest overall perceptual and paired reconstruction performance while maintaining strong preservation of the historical input.
The results also reveal a quality--preservation trade-off that is not fully captured by any single objective metric. In future work, we plan on minimizing hallucinations and extending restoration beyond orchestral classical music.
\section{Acknowledgments}
We acknowledge the use of coding agents for the implementation of experiments of this paper.


\newpage
\bibliographystyle{IEEEbib}
\bibliography{main}


\end{document}